\documentclass[pra,twocolumn,notitlepage,superscriptaddress,showpacs, amsmath, amssymb]{revtex4-1}
\usepackage{graphicx}
\usepackage{bm}
\usepackage{amsfonts}
\usepackage{latexsym}

\usepackage[usenames]{color}
\usepackage{ulem}

\newcommand{\nc}{\newcommand}
\nc{\bra}[1]{\langle #1|}
\nc{\ket}[1]{|#1\rangle}
\nc{\braket}[1]{\left\langle #1 \right\rangle}
\nc{\equ}[1]{\begin{eqnarray*}#1\end{eqnarray*}}
\nc{\equn}[1]{\begin{eqnarray}#1\end{eqnarray}}
\nc{\dagg}{^{\dagger}}
\nc{\conj}{^{*}}
\nc{\dx}[1]{\, \mathrm{d} {#1} \,}
\nc{\Dx}[1]{\mathcal{D} {#1} \,}
\nc{\la}{\langle}
\nc{\ra}{\rangle}
\nc{\tr}{\text{tr}}
\nc{\Tr}{\text{Tr} \,}
\nc{\e}{\text{e}}
\nc{\Id}{\mathbb{1}}
\nc{\eps}{\epsilon}
\nc{\der}[2]{\frac{\mathrm{d} {#1}}{\mathrm{d} {#2}}}
\nc{\pder}[2]{\frac{\partial {#1}}{\partial {#2}}}
\nc{\bigO}{\mathcal{O}}
\nc{\half}{\frac{1}{2}}

\nc{\DelA}{\Delta_{\rm A}}
\nc{\DelB}{\Delta_{\rm B}}
\nc{\delA}{\delta_{\rm A}}
\nc{\delB}{\delta_{\rm B}}
\nc{\alA}{\alpha_{\rm A}}
\nc{\nA}{n_{\rm A}}
\nc{\epsA}{\epsilon_{\rm A}}
\nc{\epsB}{\epsilon_{\rm B}}
\nc{\kex}{\kappa_{\rm ex}}
\nc{\ki}{\kappa_{\rm i}}

\newcommand{\gamp}{\gamma_\perp}
\newcommand{\Ep}{{\cal E}_{\rm p}}
\newcommand{\Delc}{\Delta_{\rm c}}
\newcommand{\Dela}{\Delta_{\rm a}}
\newcommand{\tilv}{\tilde{v}}
\newcommand{\kapd}{\kappa_d}

\nc{\wcav}{\omega_{\rm c}}
\nc{\Vm}{V_{\rm m}}

\begin{document}
\title{Observation of dressed states of distant atoms with delocalized photons in coupled-cavities quantum electrodynamics}
\author{Shinya Kato}
\affiliation{Department of Applied Physics, Waseda University, 3-4-1 Okubo, Shinjuku, Tokyo 169-8555, Japan}
\affiliation{JST, PRESTO, 4-1-8 Honcho, Kawaguchi, Saitama, 332-0012, Japan}
\author{Nikolett N\'{e}met}
\affiliation{The Dodd-Walls Centre for Photonic and Quantum Technologies, New Zealand}
\affiliation{Department of Physics, University of Auckland, Auckland 1010, New Zealand}
\author{Kohei Senga}
\affiliation{Department of Applied Physics, Waseda University, 3-4-1 Okubo, Shinjuku, Tokyo 169-8555, Japan}
\author{Shota Mizukami}
\affiliation{Department of Applied Physics, Waseda University, 3-4-1 Okubo, Shinjuku, Tokyo 169-8555, Japan}
\author{Xinhe Huang}
\affiliation{Department of Applied Physics, Waseda University, 3-4-1 Okubo, Shinjuku, Tokyo 169-8555, Japan}
\author{Scott Parkins}
\affiliation{The Dodd-Walls Centre for Photonic and Quantum Technologies, New Zealand}
\affiliation{Department of Physics, University of Auckland, Auckland 1010, New Zealand}
\author{Takao Aoki}
\email{takao@waseda.jp}
\affiliation{Department of Applied Physics, Waseda University, 3-4-1 Okubo, Shinjuku, Tokyo 169-8555, Japan}

\begin{abstract}
In a cavity quantum electrodynamics (QED) system, where atoms coherently interact with photons in a cavity, the eigenstates of the system are the superposition states of atoms and cavity photons, the so-called dressed states of atoms. 
When two cavities are connected by an optical fiber with negligible loss, the coherent coupling between the cavities gives rise to photonic normal modes. 
One of these normal modes is the fiber-dark mode, in which photons are delocalized in the two distant cavities.
Here we demonstrate the setting of coupled-cavities QED, where two nanofiber cavity-QED systems are coherently connected by a meter-long low-loss channel in an all-fiber fashion. 
Specifically, we observe dressed states of distant atoms with delocalized photons of the fiber-dark normal mode.
Our system will provide a platform for the study of delocalized atomic and photonic states, photonic many-body physics, and distributed quantum computation.
\end{abstract}
    
    
\maketitle

%

When atoms are coherently coupled to each other via their interaction with a common mode of the electromagnetic field, they form collective states, the dynamics of which can be drastically different from that of independent atoms. For the case of atoms in free space, collective effects become observable, as a change in radiative decay rates, when inter-atomic separations are smaller than a few wavelengths \cite{Dicke_PR_1954, DeVoe_PRL_1996, Scully_Science_2009, Guerin_PRL_2016}.
Recently, such super-radiance and sub-radiance phenomena have been observed for atoms interacting with a common guided mode of a photonic crystal waveguide \cite{Goban_PRL_2015} and an optical nanofiber\cite{Solano_Nature_Commun_2017}, where atoms are separated by a macroscopic distance up to several hundred microns, much larger than the wavelength.
On the other hand, in the setting of cavity quantum electrodynamics (QED), where coherent atom-atom coupling is mediated by the confined mode of the cavity, the collective effects can be observed as coherent, reversible dynamics of the system, which is evident as a structural change in the energy spectra, in contrast to changes in atomic dissipation rates as for the above cases. For example, the vacuum Rabi splitting for the dressed states of an atom in the Jaynes-Cummings model \cite{Jaynes_Proc_IEEE_1963, Thompson_PRL_1992, Boca_PRL_2004, Maunz_PRL_2004} is enhanced by a factor of $\sqrt{N}$ for the collective dressed states in the $N$-atom Tavis-Cummings model \cite{Tavis_PR_1968, Kaluzny_PRL_1983, Raizen_PRL_1989, Ruddell_Optica_2017}.
These coherent collective effects may be extended to a configuration of coupled, but distant, cavity QED systems. In particular, when two cavities are connected via a channel whose loss is negligible compared to the coupling between each cavity and the channel, photons can deterministically propagate back and forth between the cavities many times before being lost. Such coherent coupling between two cavities gives rise to normal modes, or superpositions in certain combinations of the two cavities and the connecting fiber, each of which extends to the whole system nonlocally. Notably, one of these normal modes is a superposition of the two cavity modes but has no contribution from the connecting fiber (fiber-dark mode) \cite{Serafini_PRL_2006}. 

Here, we demonstrate an all-fiber, coupled-cavities QED system in which either a single ensemble or two distantly separated ensembles of several tens of atoms coherently interact with the delocalized normal modes of coherently coupled, distant cavities, and we observe collective dressed states of atoms with the fiber-dark mode. 
This is the demonstration of coherent, reversible coupling between distant atoms, with a separation of the order of a meter, which is made possible by the all-fiber nature of the connection between the two cavities, with a loss as low as 2\%.
Our achievement is an important step towards the physical implementation of cavity QED-based distributed quantum computation \cite{Serafini_PRL_2006, Yang_NJP_2013, Koshino_PRAppl_2017, Cao_SciRep_2018} and a quantum network \cite{Kimble_Nature_2008, Reiserer_RMP_2015}, where a large number of cavity QED systems are coherently connected by low-loss fiber channels. In such systems, quantum entanglement over the whole network can be created deterministically \cite{Pellizzari_PRL_1995, Cirac_PRL_1997}, instead of probabilistically \cite{Ritter_Nature_2012}. Our achievement also paves the way for the study of many-body physics with atoms and photons in a coupled-cavities QED system, such as quantum phase transitions of light \cite{Torma_PRL_1998, Hartmann_NaturePhys_2006, Greentree_NaturePhys_2006, Rossini_PRL_2007, Angelakis_PRA_2007, Cho_PRL_2008, Irish_PRA_2008, Hartmann_LPR_2008}.


\begin{figure*}[htbp]
\begin{center}
\includegraphics[width=\textwidth, bb=0 0 855 275]{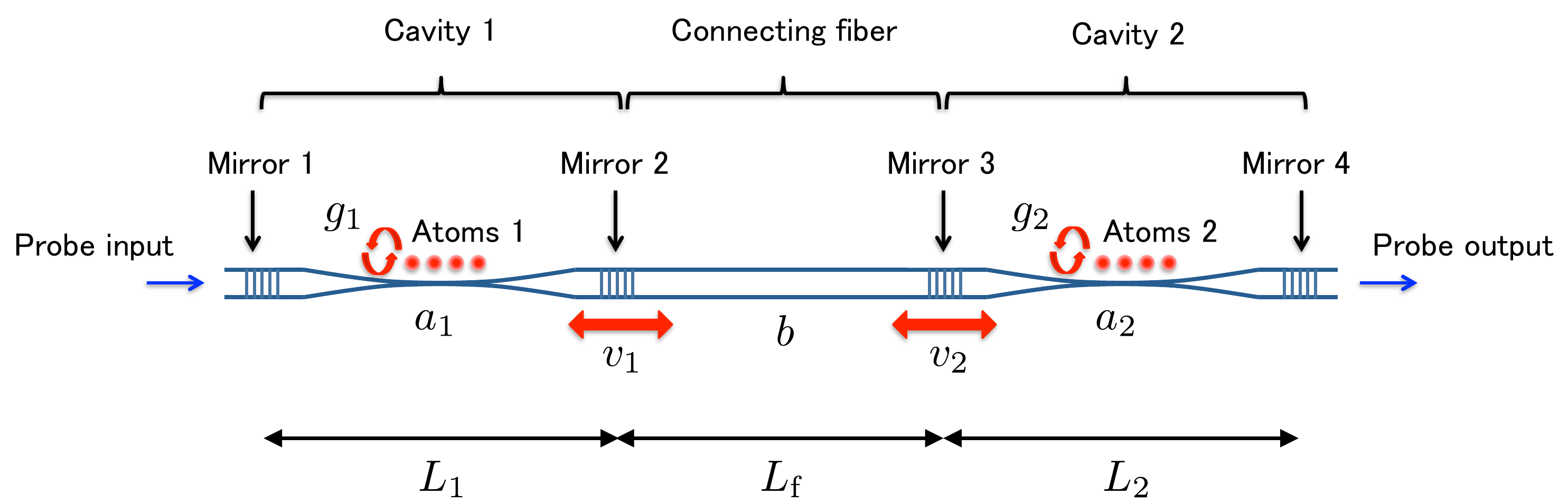}
\caption{Schematic of the coupled-cavity QED system. Cavity 1 of length $L_1$ and Cavity 2 of length $L_2$ are coupled to a connecting fiber of length $L_{\rm f}$ with coupling rates of $v_1$ and $v_2$, respectively. Atomic ensembles are coupled to Cavities 1 and 2 with atom-cavity coupling rates of $g_1$ and $g_2$, respectively. Probe beam is input from the left mirror of Cavity 1 (Mirror 1), and its output from the right mirror of Cavity 2 (Mirror 4) is measured.}
\label{fig-1}
\end{center}
\end{figure*}

As illustrated in Fig.~\ref{fig-1}, our system consists of two nanofiber cavity QED systems \cite{Kato_PRL_2015} connected in an all-fiber fashion. 
In each cavity, an ensemble of several tens of atoms interacts with the cavity field through the evanescent field of a nanofiber, both ends of which are connected to standard optical fibers through tapered regions and sandwiched by a pair of fiber-Bragg-grating mirrors.
A probe laser at the frequency  $\omega_{\rm p}$ is input from mirror 1 (left mirror of cavity 1), and the output from mirror 4 (right mirror of cavity 2) is measured.

For simplicity, we first consider the system with one atom for each cavity, which is modeled by the following Hamiltonian ($\hbar = 1$) in a frame rotating at $\omega_{\rm p}$:
\begin{eqnarray}
H &=& 
\Delta_{\rm c} \left( a_1^\dag a_1 + a_2^\dag a_2 + b^\dag b \right)
+\sum_{i=1,2} v_i \left( a_i^\dag b + b^\dag a_i \right)
\nonumber \\
&&
+\Delta_{\rm a} \left( \sigma_1^+ \sigma_1^- +  \sigma_2^+ \sigma_2^- \right)
+\sum_{i=1,2} g_i \left(a_i^\dag \sigma_i^- + \sigma_i^+ a_i \right),
\nonumber \\
&&
\end{eqnarray}
where we assume, for simplicity, that the cavity and connecting-fiber modes $(a_1, a_2, b)$ have the same frequency $\omega_{\rm c}$ so that $\Delta_{\rm c} = \omega_{\rm c} - \omega_{\rm p}$ (see Appendix). The atom-probe detuning is given by $\Delta_{\rm a} = \omega_{\rm a} - \omega_{\rm p}$, where $\omega_{\rm a}$ is the atomic transition frequency. The coupling rates of cavities 1 and 2 with the connecting fiber are given by 
\begin{eqnarray}
v_{1,2} = \frac{c}{2}\sqrt{\frac{T_{2,3}}{L_{\rm f}L_{1,2}}},
\end{eqnarray}
where $c$ is the speed of light in the fiber and $T_i$, $L_i$, and $L_{\rm f}$ are the transmittance of the mirror $i$, length of the cavity $i$, and length of the connecting fiber, respectively.
The atoms are coupled to their respective cavity modes with strengths $g_1$ and $g_2$.

Considering just the cavities and connecting fiber, we can move to a normal-mode picture, with normal-mode operators given by
\begin{eqnarray}
d &=& \frac{1}{\sqrt{2}\tilde{v}}(v_2 a_1 + v_1 a_2), \\
c_\pm &=& \frac{1}{2\tilde{v}} (v_1 a_1 + v_2 a_2) \pm \frac{1}{\sqrt{2}}b,
\end{eqnarray}
where
\begin{eqnarray}
\tilde{v} = \sqrt{\frac{v_1^2+v_2^2}{2}}.
\end{eqnarray}
Note that the normal mode $d$ has no contribution from the connecting fiber mode $b$. Rather, it has only contributions from the two cavity modes $a_1$ and $a_2$, and it has the frequency $\omega_{\rm c}$. On the other hand, the normal modes $c_\pm$ have contributions from the two cavities and the connecting fiber, and they are shifted in frequency by $\pm \sqrt{2} \tilde{v}$ from $\omega_{\rm c}$. 
If $\sqrt{2} \tilde{v}$ is sufficiently large, the atomic transition frequency is close to the bare cavity resonance, $\omega_{\rm a} \simeq \omega_{\rm c}$, and the frequency of the probe laser is scanned only in this vicinity, then it is possible to focus on the system dynamics involving only the mode $d$. That is, we can focus on the reduced Hamiltonian
\begin{eqnarray}
H_d &=& \Delta_{\rm c} d^\dag d + \sum_{i=1,2} \left[
\Delta_{\rm a} \sigma_i^+ \sigma_i^- 
+ g_{di} \left( d^\dag \sigma_i^- + \sigma_i^ + d \right) 
\right],
\nonumber \\
&&
\end{eqnarray}
where
\begin{eqnarray}
g_{d1,d2} = \frac{v_{2,1}}{\sqrt{2}\tilde{v}}g_{1,2}.
\end{eqnarray}
This Hamiltonian is identical to the Hamiltonian for a standard single-cavity QED system with a cavity mode $d$ and two atoms (Tavis-Cummings Hamiltonian\cite{Tavis_PR_1968}) having single-atom coupling strengths of $g_{d1}$ and $g_{d2}$.
The eigenstates of this system are the collective dressed states of atoms in distant cavities and of photons in the delocalized normal mode $d$. In other words, each atom interacts with both cavities simultaneously and collectively, but not with the connecting fiber. 
Indeed, the atom-field coupling strengths, $g_{d1}$ and $g_{d2}$, do not depend on the length of the connecting fiber. 

For a general case of the system with many atoms in each cavity, the linear optical response in the weak-driving limit is identical to the single-atom model discussed above, with replacements of single-atom coupling strengths $g_i$ with the collective coupling strengths $g_{i, {\rm eff}} = g_{i, (0)} \sqrt{N_{i, {\rm eff}}}$, where $g_{i, (0)}$ is the single-atom coupling strength for an atom located at a potential minimum of the atomic trap in cavity $i$ and $N_{i, {\rm eff}}$ is the effective number of atoms in cavity $i$ (see Appendix).
    

In order to investigate the interaction between atoms and the normal modes of the coupled-cavities system, we measure transmission spectra with different atom-loading conditions and various lengths of the connecting fiber. 

Each cavity QED system is similar to the previous setup described in \cite{Kato_PRL_2015}.
The transmittances of the mirrors are $(T_1, T_2, T_3, T_4) = (0.13, 0.39, 0.33, 0.06)$.
The single-pass losses inside the cavities and that for the connecting fiber are all 0.02, which are dominated by the splicing losses (two splices for each cavity and the connecting fiber). 
The cavity lengths are $(L_1, L_2) = (0.92, 1.38)$~m. 

Each experimental sequence starts with cooling and collecting Cs atoms in standard magneto-optical traps (MOTs).
We use the D$_2$-line F=4$\rightarrow$F$^\prime$=5 transition for cooling and the F=3$\rightarrow$F$^\prime$=4 transition for repumping in the MOT.
The numbers of atoms in the MOTs are 7$\times 10^6$ and 3.3$\times 10^7$ for cavities 1 and 2, respectively.
The positions of the MOTs are intentionally shifted from the cavities, and atoms are not coupled to the cavities at this stage.
After the numbers of atoms in the MOTs are saturated, we scan the lengths of the cavities and the connecting fiber with different frequencies and monitor the transmission of a laser with the frequency $\omega_{\rm a}$ (atomic F=4$\rightarrow$F$^\prime$=5 transition). 
The transmission varies with the lengths of the cavities and the connecting fiber, {\it i.e.}, the detunings of the modes $a_1, a_2,$ and $b$ from $\omega_{\rm a}$, and takes the maximum value when all the three modes are resonant to $\omega_{\rm a}$.
When the transmission reaches a certain threshold value, we assume that this resonance condition is satisfied and proceed to the next step, in which we switch off the monitor laser and change the detuning and intensity of the cooling laser for 32~ms to further cool the atoms down to 20~$\mu$K. Subsequently, we move the MOTs to overlap with the cavities to load atoms in the optical traps in the evanescent fields of the nanofibers.

Atoms are trapped in a state-insensitive, two-color evanescent-field optical trap \cite{Kien2005, Vetsch2010, Lacroute2012, Goban2012, Kato_PRL_2015}.
We use counterpropagating red-detuned beams ($\lambda_{\rm red}=$937\ nm), which form a 1-dimensional optical lattice, and a blue-detuned beam ($\lambda_{\rm blue}=$688\ nm).
The trap depth of the lattice well is about 260\ $\mu$K.

After loading atoms in the optical traps, we measure the transmission spectrum of the system by sending a probe pulse with the atom-probe detuning $\Delta_{\rm a}$ scanned over $\pm$30~MHz within 4~ms and by detecting the transmitted probe beam from the system by an avalanche photodiode after removing unwanted stray light with filters.
The power of the input probe is 210--310~pW.
Next, we optically pump the atoms into the F=3 state, which is sufficiently off-resonant from the cavity modes, by irradiating a pumping laser resonant to the atomic F=4$\rightarrow$F$^\prime$=3 transition from the sides of the cavities for 1~ms, and we send a frequency-scanned probe pulse again to measure the transmission spectrum of the coupled empty cavities, {\it i.e.}, the system without atoms.
We then switch on the cooling and repumping lasers for 4~ms to cool and load the atoms into the optical trap again. 
We repeat the above procedure for spectroscopy and re-cooling five times per MOT loading sequence, and we use the average of the second, third, and fourth data for each sequence (see Appendix for details).

\begin{figure}[htbp]
\begin{center}
\includegraphics[width=0.5\textwidth, bb=0 0 279 540]{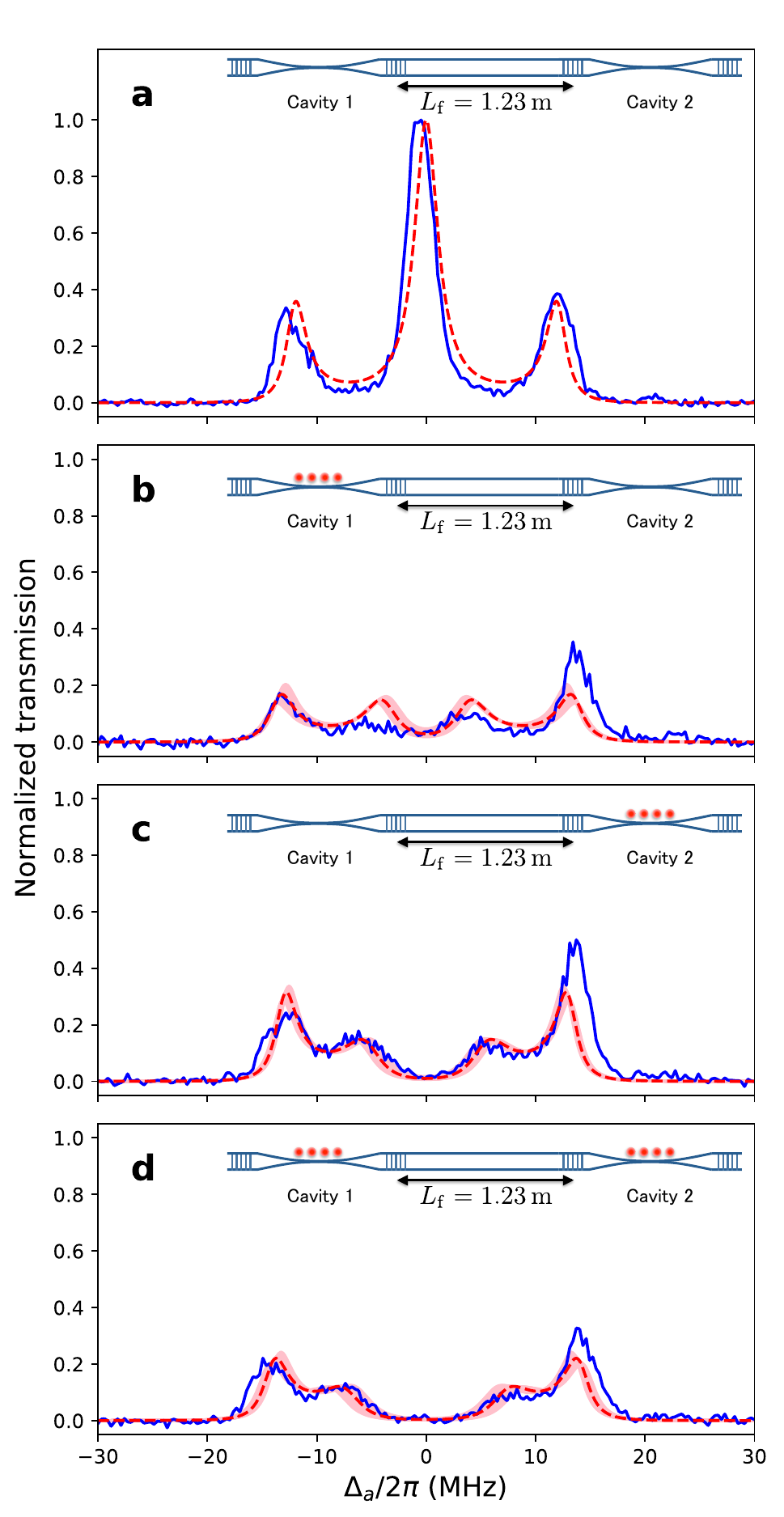}
\caption{Transmission spectra with different atom-loading conditions.
{\bf a}, No atoms are loaded. {\bf b}, Atoms are loaded in cavity 1 only.  {\bf c}, Atoms are loaded in cavity 2 only.  {\bf d}, Atoms are loaded both cavities.
Blue solid lines represent experimental data, while red dashed lines and pink shaded bands are theoretical curves for $(g_{1, {\rm eff}}, g_{2, {\rm eff}}) = 2\pi \times (7.2, 7.3)$~MHz and $(g_{1, {\rm eff}}, g_{2, {\rm eff}}) = 2\pi \times (7.2 \pm 1.0, 7.3 \pm 1.0)$~MHz, respectively. 
When no atoms are loaded ({\bf a}), formation of normal modes for three coupled empty cavities results in the clear triplet spectrum. The central peak corresponds to the fiber-dark mode $d$, while the two side peaks correspond to the other two normal modes $c_\pm$. When atoms are loaded in one of the two ({\bf b} and {\bf c}), or both cavities ({\bf d}), coupling between atoms and the fiber-dark mode results in the splitting of the central peak observed in {\bf a}. In particular, the splitting observed in {\bf d} is the signature of the dressed states of distant atoms with delocalized photons in the fiber-dark mode.
}
\label{fig-2}
\end{center}
\end{figure}

Firstly, we fix the length of the connecting fiber as $L_{\rm f} = 1.23$~m and measure transmission spectra of the coupled-cavities QED system with different atom-loading conditions, in which the atoms are not loaded, loaded in either cavity, or loaded in both cavities. 
In Fig. 2a, the blue solid line shows the measured transmission spectrum normalized to the maximum transmission for the case without atoms. 
Three peaks for the normal modes $d$ and $c_\pm$ are clearly observed. 
The red dashed line shows the corresponding theoretical curve (see Appendix) with no free fitting parameter, which agrees well with the experimental data.
In particular, the splittings of the side peaks from the center peak match with the theoretical value of the frequency difference of the modes $c_\pm$ from the fiber-dark mode $d$ given by $\sqrt{2} \tilde{v} = 2\pi \times 12.1$~MHz.
The slight broadening in linewidths and the asymmetry of the spectral shape in the experiments are, we believe, due to the instability of the detunings of the cavity and fiber modes ($a_1, a_2,$ and $b$) from the atomic frequency $\omega_{\rm a}$ during the measurement.
Figures 2b and c show the spectra for the cases of atoms loaded only in cavities 1 and 2, respectively. 
It can be clearly seen that the interaction of atoms with the fiber-dark mode $d$ causes a splitting of the center peak in Fig. 2a, which is the signature of the dressed states of atoms with delocalized photons of the fiber-dark mode. 
On the other hand, the (off-resonant) interaction of atoms with the modes $c_\pm$ causes the frequency shifts of the side peaks.
It can be seen that the spectra agree reasonably well with the theoretical curves of a linearized model with the atom-cavity coupling strengths
$(g_{1, {\rm eff}}, g_{2, {\rm eff}}) = 2\pi \times (7.2 \pm 1.0, 7.3 \pm 1.0)$~MHz ($(g_{d1}, g_{d2}) = 2\pi \times (4.3 \pm 0.6, 5.8 \pm 0.8)$~MHz)
as the only free parameters.
Figure 2d shows the spectrum for the case of atoms loaded in both cavities. 
The measured spectra clearly also agree quite well with the theoretical curves with no additional free parameters. 
Note that the splitting of the center peak is larger than that observed in Figs. 2b and c and agrees with the theoretical value of $\sqrt{g_{d1}^2 + g_{d2}^2} = 2\pi \times 7.3$~MHz, which is the signature of the collective dressed states of distant atoms with delocalized photons of the fiber-dark mode.

\begin{figure}[htbp]
\begin{center}
\includegraphics[width=0.5\textwidth, bb=0 0 279 540]{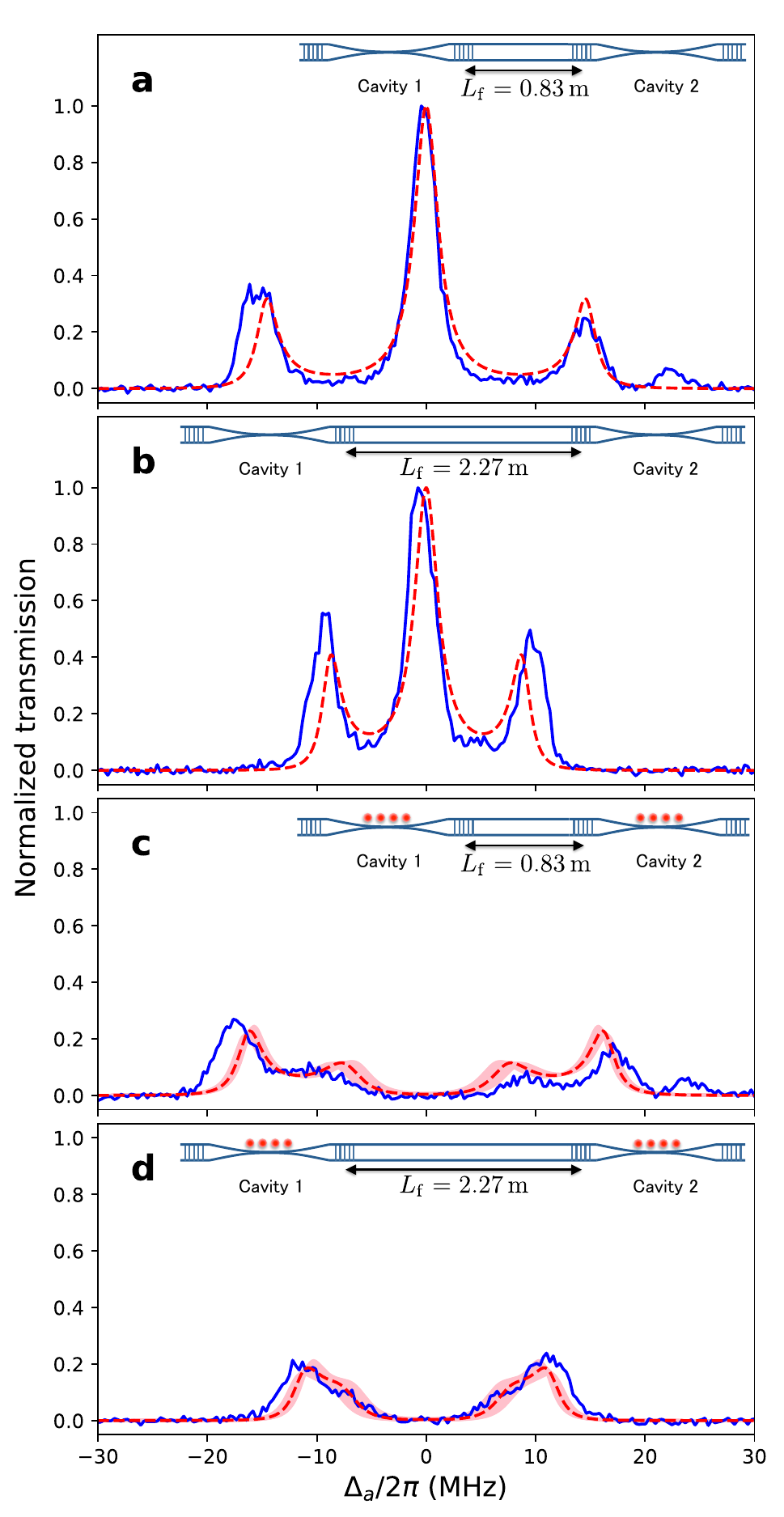}
\caption{Transmission spectra with different connecting-fiber lengths. 
Blue solid lines represent experimental data, while red dashed lines and pink shaded bands are theoretical curves for $(g_{1, {\rm eff}}, g_{2, {\rm eff}}) = 2\pi \times (7.2, 7.3)$~MHz and $(g_{1, {\rm eff}}, g_{2, {\rm eff}}) = 2\pi \times (7.2 \pm 1.0, 7.3 \pm 1.0)$~MHz, respectively.
The small peaks at $\Delta_{\rm a} \approx 2\pi \times 25$~MHz in {\bf a} and {\bf c} correspond to a mode with different polarization, which appears because of imperfect polarization compensation in the connecting fiber (see Appendix). 
The dependence of the cavity-fiber coupling rates $v_1$ and $v_2$ on the connecting-fiber length is clearly observed as the change of the splitting in the triplet structure in {\bf a} and {\bf b}. In contrast, the splitting of the central peak in {\bf c} and {\bf d} remains unchanged. This is because the coupling rate between atoms and the delocalized photons in the fiber-dark mode does not depend on the connecting-fiber length.
}
\label{fig-3}
\end{center}
\end{figure}

Secondly, we measure transmission spectra of the coupled-cavities QED system with different lengths of the connecting fiber.
Figures 3a and b show the spectra without atoms for the cases of $L_{\rm f} = 0.83$~m and 2.27~m, respectively.
The measured spectra (blue solid lines) reasonably agree with the theoretical curves (red dashed lines).
A change of splittings between the center and side peaks from that in Fig. 2a ($L_{\rm f} = 1.23$~m) is clearly observed. 
Specifically, the observed splittings roughly match the theoretical values of $\sqrt{2} \tilde{v} = 2\pi \times 14.7$~MHz and $2\pi \times$ 8.9~MHz for $L_{\rm f} = 0.83$~m and 2.27~m, respectively.
Figures 3c and d show the spectra with atoms for the cases of $L_{\rm f} = 0.83$~m and 2.27~m, respectively.
Again, the measured spectra (blue solid lines) reasonably agree with the theoretical curves (red dashed lines).
Note that the splittings of the center peak associated with the coupling of atoms to the fiber-dark mode do not change for different lengths of the connecting fiber, in agreement with the above theory.

\begin{figure}[htbp]
\begin{center}
\includegraphics[width=0.5\textwidth, bb=0 0 402 407]{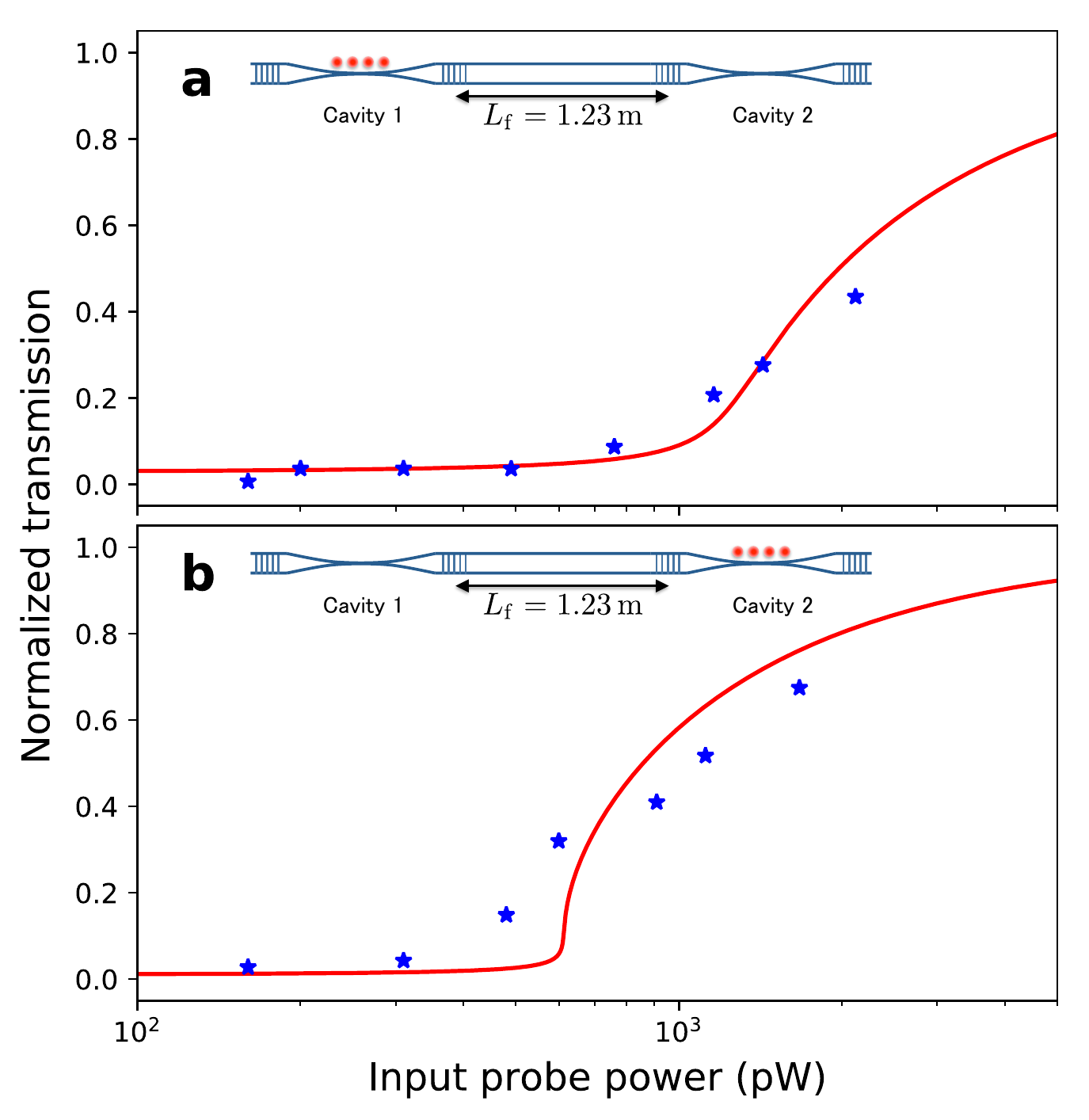}
\caption{Saturation behavior of the on-resonance transmission. Normalized transmission is plotted as a function of input probe power for atoms loaded in {\bf a}, cavity 1 and {\bf b}, cavity 2. 
Red curves are semiclassical state equations with $(g_{1, (0)}, g_{2, (0)}) = 2\pi \times (0.75, 1.2)$~MHz and $(N_{1, {\rm eff}}, N_{2, {\rm eff}}) = (92, 37)$. 
The corresponding saturation photon numbers are $(n_{1,{\rm sat}}, n_{2, {\rm sat}} ) = (6.9, 2.7)$, and thus, our system shows strong nonlinearity at the few-photon level.
}
\label{fig-4}
\end{center}
\end{figure}

Lastly, we investigate saturation behaviors of the coupled-cavities QED system to confirm that the above measurements are conducted in the weak-driving regime. 
Specifically, we load atoms in either of the cavities and measure transmission as a function of the input probe power. 
The length of the connecting fiber is fixed at $L_{\rm f} = 1.23$~m.
Figures 4a and b show the normalized transmission at zero detuning $\Delta_{\rm a} = 0$ for atoms loaded in cavities 1 and 2, respectively. 
It can be clearly seen that the system is in the weak-driving regime at the input power of 210--310~pW for the above measurement and that the transmission starts to saturate as the input power exceeds $\sim 10^3$~pW. 

We further compare the measured saturation behaviors with semiclassical state equations for the many-atom coupled-cavities QED system (see Appendix). 
The red solid lines in Figs. 4a and b are the theoretical curves with $(g_{1, (0)}, g_{2, (0)}) = 2\pi \times (0.75, 1.2)$~MHz, from which we obtain $(N_{1, {\rm eff}}, N_{2, {\rm eff}}) \equiv ((g_{1, {\rm eff}} / g_{1, (0)})^2, (g_{2, {\rm eff}} / g_{2, (0)})^2) = (92, 37)$.
It can also be seen that these curves reasonably agree with the experimental results. 
Saturation behavior of a cavity QED system can be characterized with the so-called saturation photon number, which is a measure of the system nonlinearity.
Note that, the corresponding saturation photon numbers are $(n_{1,{\rm sat}}, n_{2, {\rm sat}} ) = (6.9, 2.7)$ (see Appendix), and thus, our system shows strong nonlinearity at the few-photon level.


In summary, we have demonstrated the setting of a coupled-cavities QED system in which two nanofiber cavity-QED systems are coherently connected via a meter-long, low-loss fiber channel. Specifically, we have observed the collective dressed states of distant atoms with delocalized photons of a fiber-dark normal mode. 
We note that the delocalization of photons for the fiber-dark mode would be explicitly demonstrated by the inability to drive or detect this mode through the connecting fiber, which could be facilitated by inserting a fiber beam splitter in the connecting fiber.
Also, while in the present study the coherent coupling between distant atoms has been observed in the steady-state spectra, it will be possible, and very interesting, to also investigate transient dynamics of this system, where deterministic, reversible exchange of excitation between distant atoms would be observable.

It is straightforward to increase the number of coupled cavities in our system, and by driving the atoms from the side of the cavities with classical fields, it will be possible to realize a system of strongly interacting polaritons \cite{Hartmann_NaturePhys_2006}. Since the cavities are coupled via optical fibers, it is also possible to design a system with arbitrary geometry of connections. 
Note that ensembles of atoms $\gtrsim 10^3$ are routinely loaded into the optical traps around nanofibers \cite{Vetsch2010, Goban2012, Beguin2014}.
By reducing the number of atoms in each cavity to one \cite{Kato_PRL_2015}, on the other hand, it will be possible to construct a fiber network of single-atom cavity QED systems \cite{Kimble_Nature_2008, Reiserer_RMP_2015}. With such a system, deterministic creation of quantum entanglement over the whole network \cite{Pellizzari_PRL_1995, Cirac_PRL_1997}, instead of probabilistic creation \cite{Ritter_Nature_2012}, will be possible. 
Furthermore, provided that the normal mode description is valid, i.e., if the coherent coupling rate between each cavity and the fiber is larger than the loss rates, the interaction of atoms with the resulting fiber-dark mode can be utilized in quantum gates for distributed quantum
computation \cite{Serafini_PRL_2006}.
In order to extend our work to the construction of a fiber network of coherently coupled single-atom cavity QED systems, we are currently making technical improvements to the setup, for example, further reduction of the internal losses in the cavities, active stabilization of the cavities, and extending the lifetime of the atomic traps \cite{Meng_PRX_2018}.

\clearpage

\appendix

\begin{widetext}

\section{Theory}

\subsection{Coupled-cavities system}

\begin{figure}[h]
\centerline{
\includegraphics[scale=0.75, bb=0 0 546 116]{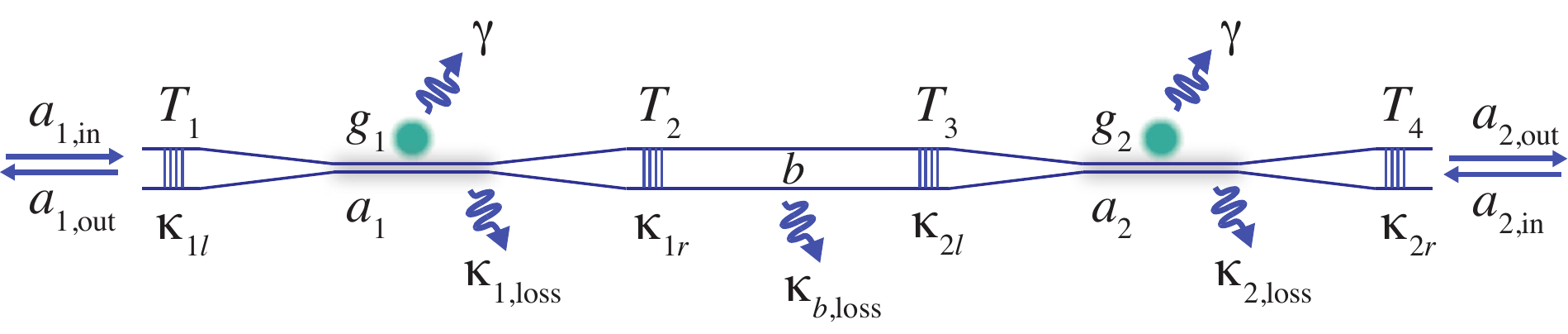}
}
\caption{Schematic of the coupled-cavities system (not to scale). The transmittance of mirror $i$ is $T_i$. Other parameters are defined in the text.}\label{fig:schematic}
\end{figure}

As a simple model of our system (Fig.~\ref{fig:schematic}), we can use single modes for the fields in the cavities and in the connecting fiber, as well as single, two-level atoms in each cavity, and a master equation for the density operator $\rho$ of the composite system (atoms plus fields) that takes the form (in a frame rotating at the probe laser frequency $\omega_{\rm p}$, and setting $\hbar=1$)
\begin{align}\label{eq:ME1}
\dot{\rho} &= -i[H,\rho ] + \kappa_1 {\cal D}[a_1]\rho + \kappa_2 {\cal D}[a_2]\rho + \kappa_{b,{\rm loss}} {\cal D}[b]\rho 
+ \frac{\gamma_\parallel}{2} \left( {\cal D}[\sigma_1^-]\rho + {\cal D}[\sigma_2^-]\rho \right) 
\nonumber
\\
& ~~~~~~~ + \gamma_{\rm las} \left( {\cal D}[a_1^\dag a_1]\rho + {\cal D}[a_2^\dag a_2]\rho + {\cal D}[b^\dag b]\rho 
+ {\cal D}[\sigma_{z,1}]\rho + {\cal D}[\sigma_{z,2}]\rho \right) .
\end{align}
where ${\cal D}[O]\rho =2O\rho O^\dag - O^\dag O\rho -\rho O^\dag O$. The Hamiltonian is
\begin{align}
H &= \Delc \left( a_1^\dag a_1 + a_2^\dag a_2 + b^\dag b \right) + \left( v_1^\ast a_1^\dag b + v_1 b^\dag a_1 \right) + \left( v_2^\ast a_2^\dag b + v_2 b^\dag a_2 \right) + \left( \Ep^\ast a_1 + \Ep a_1^\dag \right) \nonumber
\\
& ~~~ + \Dela \left( \sigma_1^+\sigma_1^- + \sigma_2^+\sigma_2^- \right) + \left( g_1 a_1^\dag \sigma_1^- + g_1^\ast \sigma_1^+ a_1 \right) + \left( g_2 a_2^\dag \sigma_2^- + g_2^\ast \sigma_2^+ a_2 \right) ,
\end{align}
where we assume, for simplicity, that the cavity and connecting-fiber modes ($a_1$, $a_2$, $b$) have the same frequency $\wcav$, so that $\Delta_{\rm c}=\wcav -\omega_{\rm p}$. The atom-probe detuning is $\Delta_{\rm a}=\omega_{\rm a}-\omega_{\rm p}$, where $\omega_{\rm a}$ is the atomic transition frequency, and $\Ep$ is the probe driving strength. The atoms couple with strengths $g_{1,2}$ to their respective cavity modes, while the coupling rates between the cavity modes (of lengths $L_{1,2}$) and the fiber mode (of length $L_{\rm f}$) are given by
\begin{align}
v_1 = \sqrt{\frac{\kappa_{1r}}{\pi} \omega_{\rm FSR,f}} \equiv \frac{c}{2}\sqrt{\frac{T_2}{L_1L_{\rm f}}} ~~~~\textrm{and}~~~~ v_2 = \sqrt{\frac{\kappa_{2l}}{\pi} \omega_{\rm FSR,f}} \equiv \frac{c}{2}\sqrt{\frac{T_3}{L_2L_{\rm f}}} .
\end{align}
Here, $\omega_{\rm FSR,f}=\pi c/L_{\rm f}$ is the free spectral range of the coupling fiber mode, where $c$ is the speed of light in the fiber, and $\kappa_{1r}=cT_2/(4L_1)$ and $\kappa_{2l}=cT_3/(4L_2)$ correspond to the decay rates of the respective cavity fields through mirrors 2 and 3 in the case that the outputs from these mirrors couple to a continuum of modes (for example, in the limit that $L_{\rm f}\rightarrow\infty$).

The remaining terms in the master equation describe losses and dephasing effects in the system. The fiber mode $b$ has an intrinsic loss rate $\kappa_{b,{\rm loss}}$, while the field decay rates of cavities 1 and 2 are given by
\begin{align}
\kappa_1 = \kappa_{1l} + \kappa_{\rm 1,loss} ~~~~\textrm{and}~~~~ \kappa_2 = \kappa_{2r} + \kappa_{\rm 2,loss} ,
\end{align}
where $\kappa_{1l}=cT_1/(4L_1)$ and $\kappa_{2r}=cT_4/(4L_2)$. 
The intrinsic loss rates are determined from the (intensity) transmission coefficients of the fiber segments that support the various modes as
\begin{align}
\kappa_{\rm 1,loss} = - \frac{1}{2} \frac{c}{L_1} \ln (1-\alpha_1) , ~~~~ \kappa_{b,{\rm loss}} = - \frac{1}{2} \frac{c}{L_{\rm f}} \ln (1-\alpha_{\rm f}) , ~~~~ \kappa_{\rm 2,loss} = - \frac{1}{2} \frac{c}{L_2} \ln (1-\alpha_2) ,
\end{align}
where $\alpha_1$, $\alpha_{\rm f}$, $\alpha_2$ are single-pass losses for the segments in cavity 1, the connecting fiber, and cavity 2, respectively.
The term proportional to $\gamma_{\rm las}$ -- the laser linewidth (HWHM) -- is included so as to incorporate the effect of laser frequency fluctuations, which appears as phase damping of the field and atomic amplitudes. The atoms decay into free space with rate $\gamma_\parallel$.

\begin{table}[h]
\begin{center}
 \begin{tabular}{|c|c|} 
 \hline
~Parameter~ & ~$2\pi\cdot$MHz~ \\ [0.5ex] 
 \hline\hline
$\kappa_{1l}$ & 1.16 \\ 
 \hline
$\kappa_{\rm 1,loss}$ & 0.36 \\
 \hline
$\kappa_{1r}$ & 3.48  \\
  \hline
$\kappa_{2l}$ & 1.97 \\ 
 \hline
$\kappa_{\rm 2,loss}$ & 0.24  \\
 \hline
$\kappa_{2r}$ & 0.357  \\ 
  \hline\hline
~$\kappa_{b,{\rm loss}}$~ ($L_{\rm f}=0.83\,{\rm m}$) & 0.40 \\ 
  \hline
$\kappa_{b,{\rm loss}}$ ($L_{\rm f}=1.23\,{\rm m}$) & 0.27  \\ 
  \hline
$\kappa_{b,{\rm loss}}$ ($L_{\rm f}=2.27\,{\rm m}$) & 0.15  \\ 
  \hline\hline
$v_1$ ($L_{\rm f}=0.83\,{\rm m}$) & 11.7 \\ 
  \hline
$v_1$ ($L_{\rm f}=1.23\,{\rm m}$) & 9.65  \\ 
  \hline
$v_1$ ($L_{\rm f}=2.27\,{\rm m}$) & 7.10  \\ 
  \hline\hline
$v_2$ ($L_{\rm f}=0.83\,{\rm m}$) & 8.82 \\ 
  \hline
$v_2$ ($L_{\rm f}=1.23\,{\rm m}$) & 7.25  \\ 
  \hline
$v_2$ ($L_{\rm f}=2.27\,{\rm m}$) &5.33  \\ 
\hline\hline
$\gamma_\parallel$ & 5.2 \\
\hline
$\gamma_{\rm las}$ & 0.365 \\
\hline
\end{tabular}
\end{center}
\caption{List of parameter values for modeling of the experiment.
$\kappa_{1l}$, $\kappa_{1r}$, $\kappa_{2l}$, $\kappa_{2r}$ correspond to the decay rates of the respective cavity fields through mirrors 1, 2, 3, and 4 (in the case that the outputs from these mirrors couple to a continuum of modes). 
$\kappa_{1,\rm loss}$, $\kappa_{b,\rm loss}$, $\kappa_{2,\rm loss}$ are the intrinsic loss rates for cavity 1, connecting fiber, and cavity 2. 
$v_1$ and $v_2$ are the coupling rates between the cavity modes and the fiber mode.
$\gamma_\parallel$ and $\gamma_{\rm las}$ are the atomic decay rate into free space and the laser linewidth (HWHM), respectively.
}
\end{table}

\subsection{Weak probe driving: linearized equations of motion}

If we assume weak driving and, hence, weak excitation of the atoms, then we may derive the following linear equations of motion for the field and atomic amplitudes:
\begin{align}\label{eq:LEOM}
\dot{\braket{a_1}} &= -(\kappa_1'+i\Delc )\braket{a_1} - iv_1\braket{b} - ig_1\braket{\sigma_1^-} - i{\cal E}_1 ,
\\
\dot{\braket{a_2}} &= -(\kappa_2'+i\Delc )\braket{a_2} - iv_2\braket{b} - ig_2\braket{\sigma_2^-}  ,
\\
\dot{\braket{b}} &= -(\kappa_b+i\Delc )\braket{b} - iv_1^\ast \braket{a_1} - iv_2^\ast \braket{a_2} ,
\\
\dot{\braket{\sigma_1^-}} &= -(\gamp+i\Dela)\braket{\sigma_1^-} - ig_1^\ast \braket{a_1}  ,
\\
\dot{\braket{\sigma_2^-}} &= -(\gamp+i\Dela)\braket{\sigma_2^-} - ig_2^\ast \braket{a_2} ,
\end{align}
where $\kappa_{1,2}'=\kappa_{1,2}+\gamma_{\rm las}$, $\kappa_b=\kappa_{b,{\rm loss}}+\gamma_{\rm las}$, and 
$\gamp=\gamma_\parallel/2+\gamma_{\rm las}$.
Setting the time derivatives to zero, we find the general steady state solution for the amplitude of cavity 2 as
\begin{align}\label{eq:a2ss}
\braket{a_2}_{\rm ss} = \frac{A}{B} \, ,
\end{align}
where
\begin{align}\label{eq:A}
A =  - i{\cal E}_1 \left( \frac{v_2}{\kappa_b+i\Delc }\right) \frac{v_1^\ast}{\kappa_1'+i\Delc +\dfrac{|g_1|^2}{\gamp+i\Dela}+\dfrac{|v_1|^2}{\kappa_b+i\Delc }} ,
\end{align}
and
\begin{align}\label{eq:B}
B = -(\kappa_2'+i\Delc ) - \frac{|v_2|^2}{\kappa_b+i\Delc } - \frac{|g_2|^2}{\gamp+i\Dela} + \frac{|v_1v_2|^2}{(\kappa_b+i\Delc )^2}\, \frac{1}{\kappa_1'+i\Delc +\dfrac{|g_1|^2}{\gamp+i\Dela}+\dfrac{|v_1|^2}{\kappa_b+i\Delc }} .
\end{align}
Solutions for the steady state amplitudes of cavity 1 and fiber mode $b$ then follow from
\begin{align}\label{eq:a1ss}
\braket{a_1}_{\rm ss} = -\frac{i{\cal E}_1+ \dfrac{v_1v_2^\ast}{\kappa_b+i\Delc }\,\braket{a_2}_{\rm ss}}{\kappa_1'+i\Delc +\dfrac{|g_1|^2}{\gamp+i\Dela}+\dfrac{|v_1|^2}{\kappa_b+i\Delc }} ,
\end{align}
and
\begin{align}\label{eq:bss}
\braket{b}_{\rm ss} = - \frac{iv_1^\ast}{\kappa_b+i\Delc }\,\braket{a_1}_{\rm ss} - \frac{iv_2^\ast}{\kappa_b+i\Delc }\,\braket{a_2}_{\rm ss}  .
\end{align}
The output photon flux from cavity 2, in the linear regime, is then  
\begin{align}
\left|\braket{a_{\rm out,2}}\right|^2 = 2\kappa_{2r} \left|\braket{a_2}\right|^2 .
\end{align}
When plotting this quantity, we normalize by the on-resonance ($\Delc =0$) flux with no atoms ($g_{1,2}=0$).

\subsection{Normal mode description}

We assume, for simplicity, that all of the parameters in the Hamiltonian are real. Considering just the cavities and coupling fiber, we can move to a normal mode picture, with normal mode operators given by
\begin{align}
d = \frac{1}{\sqrt{2}\tilv} \left( v_2 a_1 + v_1 a_2 \right) , ~~~~
c_\pm =  \frac{1}{2\tilv} \left( v_1 a_1 + v_2 a_2 \right)  \pm \frac{1}{\sqrt{2}} b ,
\end{align}
where
\begin{align}
\tilv = \sqrt{\frac{\left( v_1^2+v_2^2\right)}{2}} .
\end{align}
Expressed in terms of these normal mode operators, the Hamiltonian becomes
\begin{align}
H &= \Delc d^\dag d + \left( \Delc + \sqrt{2}\tilv \right) c_+^\dag c_+ + \left( \Delc - \sqrt{2}\tilv \right) c_-^\dag c_- + \Dela \left( \sigma_1^+\sigma_1^- + \sigma_2^+\sigma_2^- \right)  \nonumber
\\
& ~~~ + \Ep \frac{v_1}{2\tilv} \left( c_+^\dag + c_+ \right) + \Ep \frac{v_1}{2\tilv} \left( c_-^\dag + c_- \right) + \Ep \frac{v_2}{\sqrt{2}\tilv} \left( d^\dag + d \right) \nonumber
\\
& ~~~ + \frac{1}{2\tilv} \left[ \left( c_+^\dag + c_-^\dag \right) \left( v_1g_1 \sigma_1^- + v_2g_2 \sigma_2^- \right) + \textrm{H.c.} \right] + \frac{1}{\sqrt{2}\tilv} \left[ d^\dag \left( v_2g_1 \sigma_1^- + v_1g_2 \sigma_2^- \right) + \textrm{H.c.} \right] .
\end{align}
Let us now assume that the normal mode splitting $\sqrt{2}\tilv$ is large compared to all other parameters. If the atomic transition frequency is close to the bare cavity resonance ($\omega_{\rm a}\simeq\wcav$), and the probe laser frequency is scanned just in this vicinity as well, then it is possible to focus on the system dynamics involving only the resonant mode $d$. That is, we can focus on the reduced Hamiltonian
\begin{align}
H_d &= \Delc d^\dag d + \Dela \left( \sigma_1^+\sigma_1^- + \sigma_2^+\sigma_2^- \right)  
+ \Ep \frac{v_2}{\sqrt{2}\tilv} \left( d^\dag + d \right) 
+ \frac{1}{\sqrt{2}\tilv} \left[ d^\dag \left( v_2g_1 \sigma_1^- + v_1g_2 \sigma_2^- \right) + \textrm{H.c.} \right] 
\nonumber
\\
& \equiv \Delc d^\dag d + \sum_{i=1,2} \left[ \Dela \sigma_i^+\sigma_i^- + g_{di} \left( d^\dag \sigma_i^- + \sigma_i^+ d \right) \right] + {\cal E}_d \left( d^\dag + d \right) ,
\end{align}
where 
\begin{align}
g_{d1} = \frac{v_2}{\sqrt{2}\tilv} g_1 , ~~~ g_{d2} = \frac{v_1}{\sqrt{2}\tilv} g_2 , ~~~ \textrm{and} ~~~  {\cal E}_d = \Ep \frac{v_2}{\sqrt{2}\tilv} .
\end{align}

Now considering the dissipative terms in the master equation and noting the inverse relations
\begin{align}\label{eq:ba1a2}
b = \frac{1}{\sqrt{2}} \left( c_+ - c_- \right), ~~~
a_1 = \frac{1}{2\tilv} \left[ v_1 \left( c_+ + c_- \right) + \sqrt{2} v_2 d \right] , ~~~
a_2 = \frac{1}{2\tilv} \left[ v_2 \left( c_+ + c_- \right) - \sqrt{2} v_1 d \right] ,
\end{align}
one can see that, in general, the decay of each normal mode cannot be decoupled from the other normal modes, i.e., there are cross-terms between the normal mode operators arising from the Lindblad forms $\kappa_1 {\cal D}[a_1]\rho + \kappa_2 {\cal D}[a_2]\rho + \kappa_{b,{\rm loss}} {\cal D}[b]\rho$ (for simplicity, we set $\gamma_{\rm las}=0$ and ignore phase damping in this discussion). Once again, however, if $\sqrt{2}\tilv$ is sufficiently large, then the cross-terms can be neglected (as in a rotating-wave approximation) and the normal modes can be regarded as decaying according to 
\begin{align}
\kappa_d {\cal D}[d]\rho + \kappa_+ {\cal D}[c_+]\rho + \kappa_- {\cal D}[c_-]\rho ,
\end{align}
with
\begin{align}
\kapd = \frac{v_2^2\kappa_1 + v_1^2\kappa_2}{2\tilv^2} ~~~~\textrm{and} ~~~~
\kappa_+ = \kappa_- = \frac{1}{2} \left( \kappa_{b,{\rm loss}} + \frac{v_1^2\kappa_1 + v_2^2\kappa_2}{2\tilv^2} \right) .
\end{align}
So, we can in turn focus on the reduced master equation,
\begin{align}\label{eq:MEdmode}
\dot{\rho} &= -i[H_d,\rho ] + \kapd {\cal D}[d]\rho + \gamma_\perp \left( {\cal D}[\sigma_1^-]\rho + {\cal D}[\sigma_2^-]\rho \right) ,
\end{align}
for probe driving around the atom-cavity resonance $\wcav\simeq\omega_{\rm a}$ and far from the normal modes at $\wcav\pm\sqrt{2}\tilv$.

\subsection{Vacuum Rabi splitting of the fiber-dark mode $d$}

The photon flux transmitted through the coupled-cavities system is determined by the output photon flux from cavity 2, i.e., by $2\kappa_{2r}\langle a_2^\dag a_2\rangle$. Given weak probe driving strength in equation (\ref{eq:MEdmode}), and the relationship between $a_2$ and $d$ in equation (\ref{eq:ba1a2}), the transmitted spectrum will show, for sufficiently large atom-cavity coupling strengths, vacuum Rabi splitting of the fiber-dark mode $d$ given by
\begin{align}
\pm \sqrt{\left( g_{d1}^2+g_{d2}^2\right) -\frac{1}{4}(\kapd -\gamp )^2} ,
\end{align}
which is approximately $\pm \sqrt{ g_{d1}^2+g_{d2}^2}$ for the parameters of the experiment.

\subsection{Saturation}

From the semiclassical (nonlinear) equations of motion in steady state, we can eliminate the atomic variables, as well as the connecting-fiber mode amplitude, to 
obtain the following coupled equations for the amplitudes of the two cavity modes,
\begin{align}
-i\frac{\Ep}{\kappa_1'} &= \braket{a_1} \left\{ \left( 1+i\frac{\Delc}{\kappa_1'} + \frac{v_1^2/\kappa_1'}{\kappa_b+i\Delc} \right) + \left( 1-i\frac{\Dela}{\gamp} \right) \sum_{j_1} \frac{C_{1j_1}}{1+\dfrac{\Dela^2}{\gamp^2}+\dfrac{\left|\braket{a_1}\right|^2}{n_{1j_1}}} \right\} + \frac{v_1v_2/\kappa_1'}{\kappa_b+i\Delc} \braket{a_2} ,
\\
0 &= \braket{a_2} \left\{ \left( 1+i\frac{\Delc}{\kappa_2'} + \frac{v_2^2/\kappa_2'}{\kappa_b+i\Delc} \right) + \left( 1-i\frac{\Dela}{\gamp} \right) \sum_{j_2} \frac{C_{2j_2}}{1+\dfrac{\Dela^2}{\gamp^2}+\dfrac{\left|\braket{a_2}\right|^2}{n_{2j_2}}} \right\} + \frac{v_1v_2/\kappa_2'}{\kappa_b+i\Delc} \braket{a_1} ,
\end{align}
where
\begin{align}
C_{lj_l} = \frac{g_{lj_l}^2}{\kappa_l'\gamp} ~~~ \textrm{and} ~~~ n_{lj_l} = \frac{\gamp\gamma_\parallel}{4g_{lj_l}^2} ~~~~~~~ (l=1,2) .
\end{align}
Here, $g_{lj_l}$ denotes the coupling strength of atom $j_l$ to cavity mode $l$. Note that we will assume for simplicity that the parameters $\{ g_{1j_1},g_{2j_2}\}$, $\{ v_1,v_2\}$, and $\Ep$ are all real. In what follows, we will also assume resonance between atoms, field modes and driving laser field, i.e., $\Delc=\Dela=0$.

%
\subsubsection{Atoms in cavity 1 only}

Setting $C_{2j_2}=0$ (no atoms in cavity 2), we can eliminate the amplitude $\braket{a_2}$ to obtain an equation for the amplitude $\braket{a_1}$ alone, which in scaled variables takes the form
\begin{align}\label{eq:y1}
-iy_1 &= X_1 \left\{ 1 +  \frac{v_1^2/(\kappa_b\kappa_1')}{1+v_2^2/(\kappa_b\kappa_2')} + C_{\rm 1,(0)}\sum_{j_1} \frac{C_{1j_1}/C_{\rm 1,(0)}}{1+\left| X_1\right|^2 \left( g_{1j_1}/g_{\rm 1,(0)}\right)^2} \right\} ,
\end{align}
where
\begin{align}
y_1 = \frac{\Ep}{\kappa_1'\sqrt{n_{\rm 1,sat}}} , ~~~ X_1 = \frac{\braket{a_1}}{\sqrt{n_{\rm 1,sat}}} , ~~~ n_{\rm 1,sat} = \frac{\gamp\gamma_\parallel}{4g_{\rm 1,(0)}^2} , ~~~ C_{\rm 1,(0)} = \frac{g_{\rm 1,(0)}^2}{\kappa_1'\gamp} .
\end{align}
Here, $g_{\rm 1,(0)}$ is the single-atom coupling strength for an atom located at a potential minimum of the dipole trap in cavity 1.
The normalized probe transmission through the system (i.e., the normalized output intensity from cavity 2) in the semiclassical model is given by
\begin{align}
T = \left| \frac{\braket{a_2}}{\braket{a_2}_0} \right|^2 = \frac{1}{y_1^2} \left( 1 + \frac{v_1^2}{\kappa_b\kappa_1'} + \frac{v_2^2}{\kappa_b\kappa_2'} \right)^2 \frac{\left| X_1\right|^2}{\left( 1 + \dfrac{v_2^2}{\kappa_b\kappa_2'}\right)^2} ,
\end{align}
where $\braket{a_2}_0$ is the mode amplitude of cavity 2 for resonant driving and no atoms in either cavity. Finally, the input probe power can be related to $y_1$ and $n_{\rm 1,sat}$ through the relation
\begin{align}\label{eq:Pin}
P_{\rm in} = \frac{\Ep^2}{2\kappa_{1l}} \hbar\omega_{\rm p} = \frac{\Ep^2}{2\kappa_{1l}} \frac{2\pi\hbar c}{\lambda} = y_1^2 \left(  \frac{2\pi\hbar c}{\lambda} \right) \left( \frac{{\kappa_1'}^2}{2\kappa_{1l}} \right) n_{\rm 1,sat} .
\end{align}
To compare theory and experiment, we plot $T$ versus $P_{\rm in}$ as given in equation (\ref{eq:Pin}), with $n_{\rm 1,sat}$ as a scaling factor.

%
\subsubsection{Atoms in cavity 2 only}

Now setting $C_{1j_2}=1$ (no atoms in cavity 1), we can similarly eliminate the amplitude $\braket{a_1}$ to obtain an equation for the amplitude $\braket{a_2}$ alone, which in scaled variables takes the form
\begin{align}\label{eq:y2}
iy_2 &= X_2 \frac{v_1\kappa_2'}{v_2\kappa_1'} \left( 1 + \frac{\kappa_b\kappa_1'}{v_1^2} \right) \left\{ 1 + \frac{v_2^2\kappa_1'}{v_1^2\kappa_2'} \frac{1}{1+\kappa_b\kappa_1'/v_1^2} + C_{\rm 2,(0)} \sum_{j_2} \frac{C_{2j_2}/C_{\rm 2,(0)}}{1+\left| X_2 \right|^2 (g_{2j_2}/g_{\rm 2,(0)})^2} \right\} ,
\end{align}
where
\begin{align}
y_2 = \frac{\Ep}{\kappa_1'\sqrt{n_{\rm 2,sat}}} , ~~~ X_2 = \frac{\braket{a_2}}{\sqrt{n_{\rm 2,sat}}} , ~~~ n_{\rm 2,sat} = \frac{\gamp\gamma_\parallel}{4g_{\rm 2,(0)}^2} , ~~~ C_{\rm 2,(0)} = \frac{g_{\rm 2,(0)}^2}{\kappa_2'\gamp} .
\end{align}
Here, $g_{\rm 2,(0)}$ is the single-atom coupling strength for an atom located at a potential minimum of the dipole trap in cavity 2.
The normalized probe transmission through the system in this case is given by
\begin{align}
T = \left| \frac{\braket{a_2}}{\braket{a_2}_0} \right|^2 = \frac{1}{y_2^2} \left( \frac{1+v_1^2/(\kappa_b\kappa_1')+v_2^2/(\kappa_b\kappa_2')}{v_1v_2/(\kappa_b\kappa_2')} \right)^2 \left| X_2\right|^2 ,
\end{align}
and the relationship to the input power becomes
\begin{align}
P_{\rm in} = \frac{\Ep^2}{2\kappa_{1l}} \hbar\omega_{\rm p} =  y_2^2 \left(  \frac{2\pi\hbar c}{\lambda} \right) \left( \frac{{\kappa_1'}^2}{2\kappa_{1l}} \right) n_{\rm 2,sat} .
\end{align}

%
\subsubsection{Position dependence of the atom-cavity coupling}

From \cite{Lacroute2012}, we take the (square of the) atom-cavity mode coupling strength, in both cavities, to have the form ($l=1,2$)
\begin{align}\label{eq:modefn}
g_l(r,\phi ,z)^2 &= {\cal C} \left(\frac{\beta}{2q}\right)^2 \left\{ \left[ (1-s)K_0(qr)+(1+s)K_2(qr)\cos(2\phi ) \right]^2 + (1+s)^2K_2^2(qr)\sin^2(2\phi ) \right\} \cos^2(\beta z) \nonumber
\\
& + {\cal C} \left\{ K_1^2(qr)\cos^2(\phi ) \right\} \sin^2(\beta z) ,
\end{align} 
where ${\cal C}$ is a constant and we use cylindrical coordinates, with the $z$-direction along the fiber axis and $r$ ($>a$) measured from the center of the fiber. The propagation constant $\beta =7.87925\times 10^6~{\rm m}^{-1}$, and $q=\sqrt{\beta^2-n_2^2k^2}$
with $n_2=1$ and $k=2\pi /\lambda$ ($\lambda =852~{\rm nm}$). The parameter $s=-0.828$ depends on $q$, on $h=\sqrt{k^2n_1^2-\beta^2}$ ($n_1=1.4525$), and on the radius ($a$) of the nanofiber  \cite{Lacroute2012}. 
With respect to the position of a minimum of the dipole trapping potential located at $(r=r_{l,0},\phi=0,z=0)$, we find that the above function is in fact very well approximated by the simplified form
\begin{align}\label{eq:modefnsimpl}
g_l(r,\phi ,z)^2 &= g_{l,(0)}^2 \frac{1}{2} \left[ 1+A+B\cos(2\beta z) \right] \frac{e^{-2q'(r-r_{l,0})}}{r/r_{l,0}} \cos^2(\phi ) ,
\end{align}
with $A+B=1$ and $q'\simeq 1.3q$. We shall use this approximate form in the calculations that follow.

%
\subsubsection{Integral approximation to the summation over atoms}

Assuming sufficiently many atoms in each cavity, we replace the summations in (\ref{eq:y1}) and (\ref{eq:y2}) as follows:
\begin{align}
C_{l,(0)} \sum_{j_l} \frac{C_{lj_l}/C_{l,(0)}}{1+\left| X_l\right|^2(g_{lj_l}/g_{l,(0)})^2} \longrightarrow C_{l,(0)} \int_0^L \int_{-\pi}^\pi \int_a^\infty \rho_l (r,\phi,z) \frac{(g_l(r,\phi ,z)/g_{l,(0)})^2}{1+\left| X_l\right|^2(g_l(r,\phi ,z)/g_{l,(0)})^2} \, r dr d\phi dz ,
\end{align}
where $\rho_l (r,\phi,z)$ is the atom density distribution in cavity $l$.
With respect to the $z$-direction, atoms are tightly confined by a standing wave optical potential. However, this standing wave is incommensurate with the cavity mode standing wave and so, on average, the atomic distribution along the $z$-direction, with regards to the cavity mode, can be regarded as uniform. That is, we take $\rho_l (r,\phi ,z)=\rho_l (r,\phi )$, and then the integration over $z$ can be carried out exactly, reducing the above expression to
\begin{align}\label{eq:intdrdphi}
C_{l,(0)} L\frac{1}{\left| X_l\right|^2}  \int_{-\pi}^\pi \int_a^\infty \rho_l (r,\phi ) f_l(r,\phi ) rdr d\phi ,
\end{align}
where
\begin{align}
f_l(r,\phi ) =  1 - \left[ \left(1+A\left| X_l\right|^2 \dfrac{e^{-2q'(r-r_0)}}{r/r_0} \cos^2(\phi ) \right) \left(1+\left| X_l\right|^2 \dfrac{e^{-2q'(r-r_0)}}{r/r_0} \cos^2(\phi ) \right)\right]^{-1/2} .
\end{align}
Making a harmonic approximation to the atom trapping potential, the atomic density distribution can be written, in Cartesian coordinates, as
\begin{align}
\rho_l (x,y) = \rho_{l,0} e^{-(x-x_{l,0})^2/\sigma_{l,x}^2} e^{-y^2/\sigma_{l,y}^2} ,
\end{align}
where $x_{l,0}=r_{l,0}$ and $\sigma_{x,y}=\sqrt{2k_{\rm B}T/m\omega_{x,y}^2}$ for a gas at temperature $T$ and with trapping frequencies $\omega_{x,y}$. 

The double integral (\ref{eq:intdrdphi}) can be rewritten in Cartesian coordinates as
\begin{align}
C_{l,(0)} L\rho_{l,0} \frac{1}{\left| X_l\right|^2}  \iint_{x^2+y^2\geq a^2} e^{-(x-x_{l,0})^2/\sigma_{l,x}^2} e^{-y^2/\sigma_{l,y}^2} f_l(x,y) \, dx dy .
\end{align}
In the $x$-direction, the function $f_l(x,y)$ varies slowly (i.e., approximately linearly) in comparison to the Gaussian density distribution for the characteristic parameters of the experiment. Given this, and assuming $\sigma_{l,x}\ll x_{l,0}-a$ ($x_{l,0}-a$ is the distance from the trap center to the surface of the nanofiber), the integral is well approximated by
\begin{align}\label{eq:intdu}
C_{l,(0)} L\rho_{l,0} \frac{1}{\left| X_l\right|^2}  \sqrt{\pi}\sigma_{l,x} \sigma_{l,y} \int_{-\infty}^\infty e^{-u^2} f_l(x_{l,0},u)\, du ,
\end{align}
where 
\begin{align}
f_l(x_{l,0},u) = 1 - \frac{1}{\sqrt{\left( 1+A\left| X_l\right|^2 s(x_{l,0},u) \right) \left( 1+\left| X_l\right|^2 s(x_{l,0},u) \right)}} ,
\end{align}
with
\begin{align}
s(x_{l,0},u) = \frac{\exp\left[ -2q'x_{l,0}\left(\sqrt{1+(\sigma_{l,y}^2/x_{l,0}^2)u^2}-1\right)\right]}{\left[1+(\sigma_{l,y}^2/x_{l,0}^2)u^2\right]^{3/2}} .
\end{align}
The trapping along the $y$-direction is typically much weaker than in the other directions (i.e., $\sigma_{l,y}\gtrsim 5\sigma_{l,x}$), but, given uncertainty in the exact trapping parameters, and making the (not unreasonable) assumption that $\sigma_{l,y}^2/x_{l,0}^2\ll 1$, we can set 
$s(x_{l,0},u)=1$ in the integral, and then (\ref{eq:intdu}) reduces to 
\begin{align}\label{eq:planeSW}
C_{l,(0)} N_{l,{\rm eff}}
\frac{2}{(1+A)} \frac{1}{\left| X_l\right|^2} \left[ 1 - \frac{1}{\sqrt{( 1+A\left| X_l\right|^2 ) ( 1+\left| X_l\right|^2 )}} \right] ,
\end{align}
where $N_{l,{\rm eff}} = \rho_{l,0} (L/2) (1+A) \pi \sigma_{l,x}\sigma_{l,y}$. In our comparison between theory and experiment, we use (\ref{eq:planeSW}) to approximate the summations in (\ref{eq:y1}) and (\ref{eq:y2}). We use a value $A=0.17$, deduced from the experimental parameters and (\ref{eq:modefn}). As mentioned earlier, matching the theory curves for $T$ versus $P_{\rm in}$ to the experimental data yields  estimates of $n_{l,{\rm sat}}$, from which we can deduce values for $g_{l,(0)}$. From these, and the estimates of $g_{l,{\rm eff}}$ obtained from the weak-field transmission spectra, we then deduce estimates for $N_{l,{\rm eff}}$.

\end{widetext}

\section{Experimental Setup}
\begin{figure*}[htbp]
\includegraphics[width=1.0\textwidth, bb=9 9 1182 438]{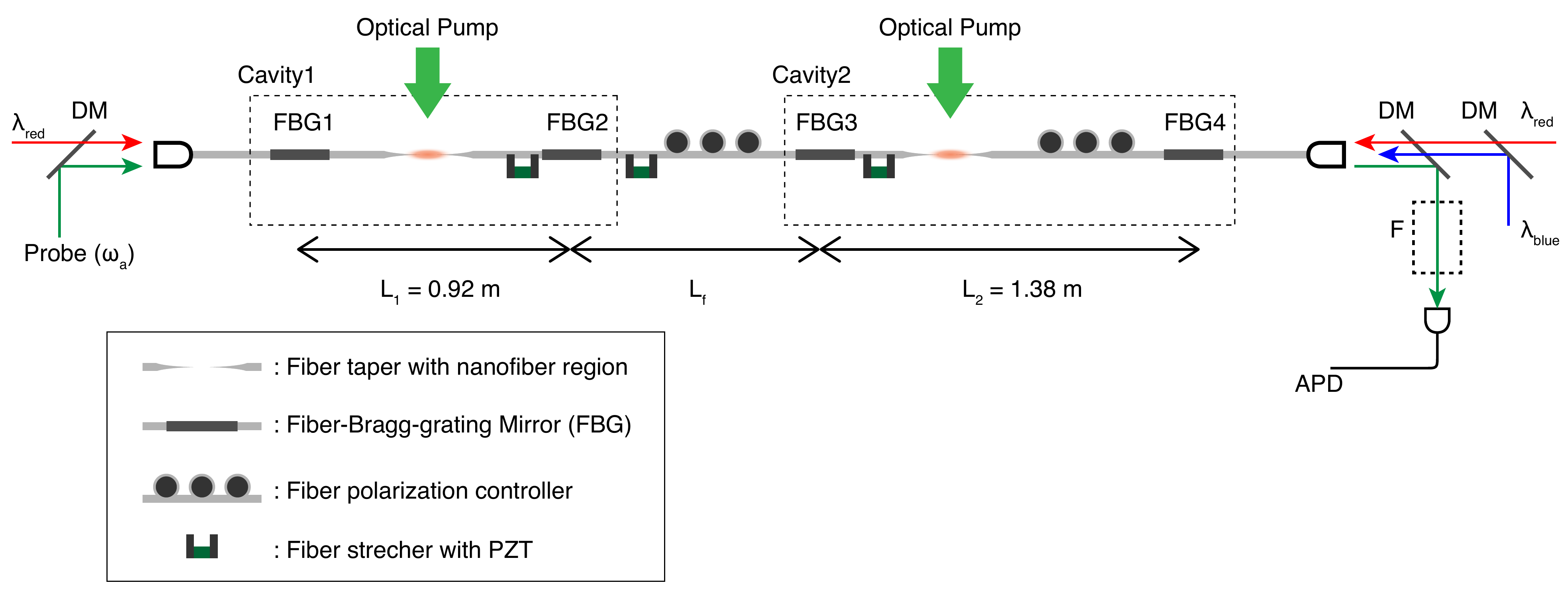}
\caption{Schematic of the setup.  Cavity 1 of length $L_1=0.92$~m and Cavity 2 of length $L_2=1.38$~m are connected by a fiber of length $L_{\rm f}$.
DM: Dichroic mirror, F: filters, APD: avalanche photodiode.
 }
\label{sfig-1}
\end{figure*}

The schematic of the experimental setup is shown in Fig.~\ref{sfig-1}.
Each cavity consists of a nanofiber, both ends of which are connected to standard single-mode fibers through tapered regions, and sandwiched by a pair of fiber-Bragg-grating (FBG) mirrors\cite{Kato_PRL_2015}.
The single-pass loss inside each cavity is 2\%, which is dominated by the losses in two fiber splices.
The diameter and length of the nanofibers are 400\ nm and 3\ mm, respectively.
The nanofibers are located in two separate UHV chambers.

We put three-paddle polarization controllers in the connecting fiber and cavity 2 to compensate polarization rotation in the optical fiber.
For the case of $L_{\rm f} = 0.83$~m (Figs. 3a and c in the main text), we replace the polarization controller in the connecting fiber with a two-paddle one because of the limitation of the fiber length, which cause the imperfect polarization compensation.
The lasers for the optical trap and probe are mixed by using dichroic mirrors and coupled to the optical fiber.
In contrast, the optical pumping pulses are irradiated on the nanofiber from its side.
The polarizations of the trapping fields are linearly polarized and parallel to each other at the nanofiber regions.
The transmitted probe beam passes the filtering system to block unwanted stray light, and is detected by an avalanche photodetector.
The detection efficiency including the transmittance of the filtering system and the quantum efficiency of the detector is 0.26.

\section{Measurement sequence and data analysis}

\begin{figure}[htbp]
\includegraphics[width=0.5\textwidth, bb=0 0 459 288]{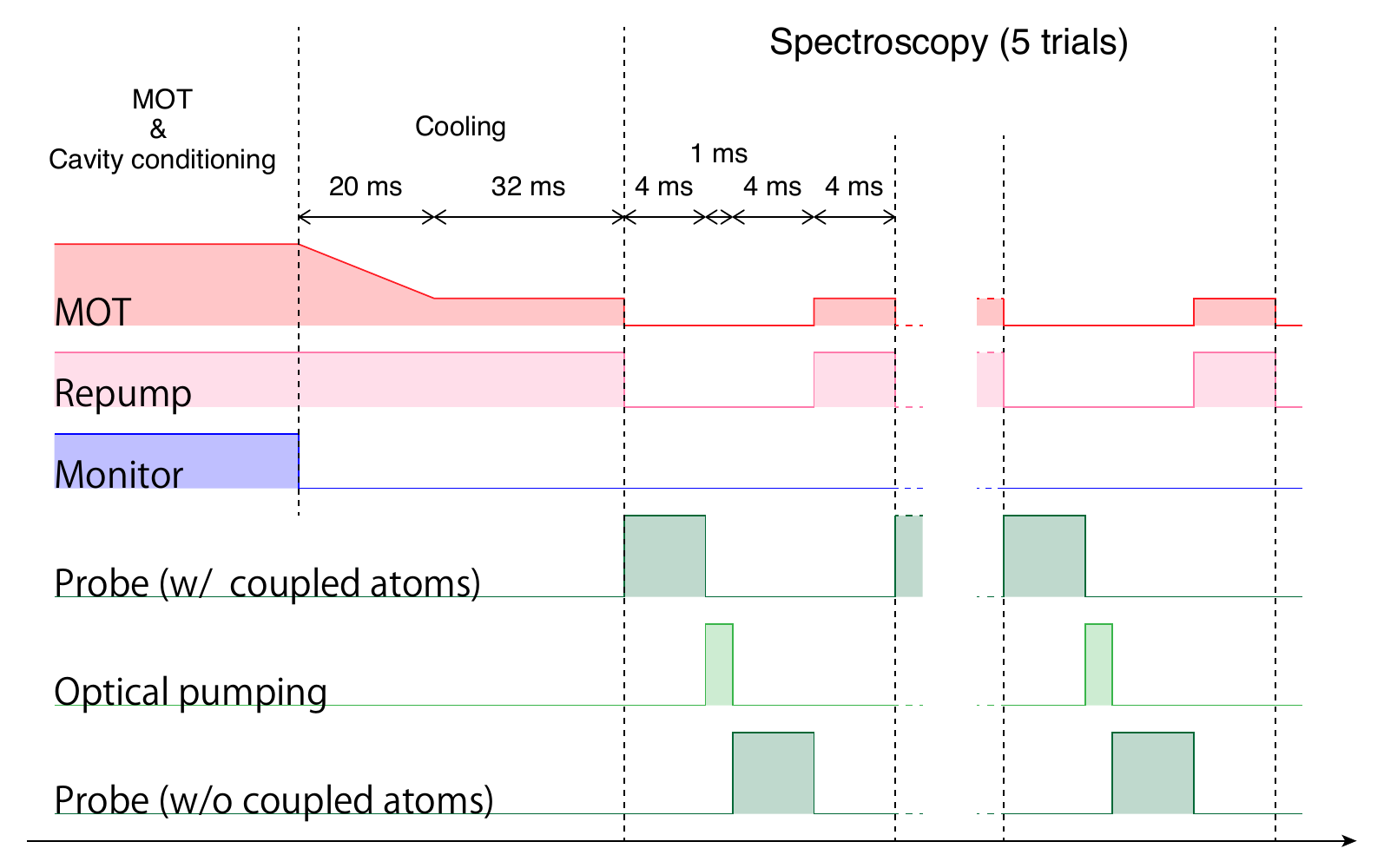}
\caption{Sequence diagram of the spectroscopy. After the preparation of atoms in the MOT and additional cooling, the spectroscopy procedure repeats five times.
}
\label{sfig-2}
\end{figure}

The schematic diagram of the pulse sequence for the transmission spectroscopy is shown in Fig.\ \ref{sfig-2}.
Each sequence starts from cooling and trapping cesium atoms in the two UHV chambers by using standard six-beam magneto-optical traps (MOT).
We use the D$_2$ line F=4$\rightarrow$F'=5 transition for cooling and the F=3$\rightarrow$F'=4 transition for repumping in the MOTs.
In the loading stage of the MOTs, the total intensity and the detuning of the cooling beams are, respectively, 16$I_s$ and -1.25$\Gamma$, where $I_s$ and $\Gamma$ are the saturation intensity and natural linewidth of the cooling transition.

\begin{figure*}[htbp]
\includegraphics[width=0.4\textwidth, bb=0 0 432 864]{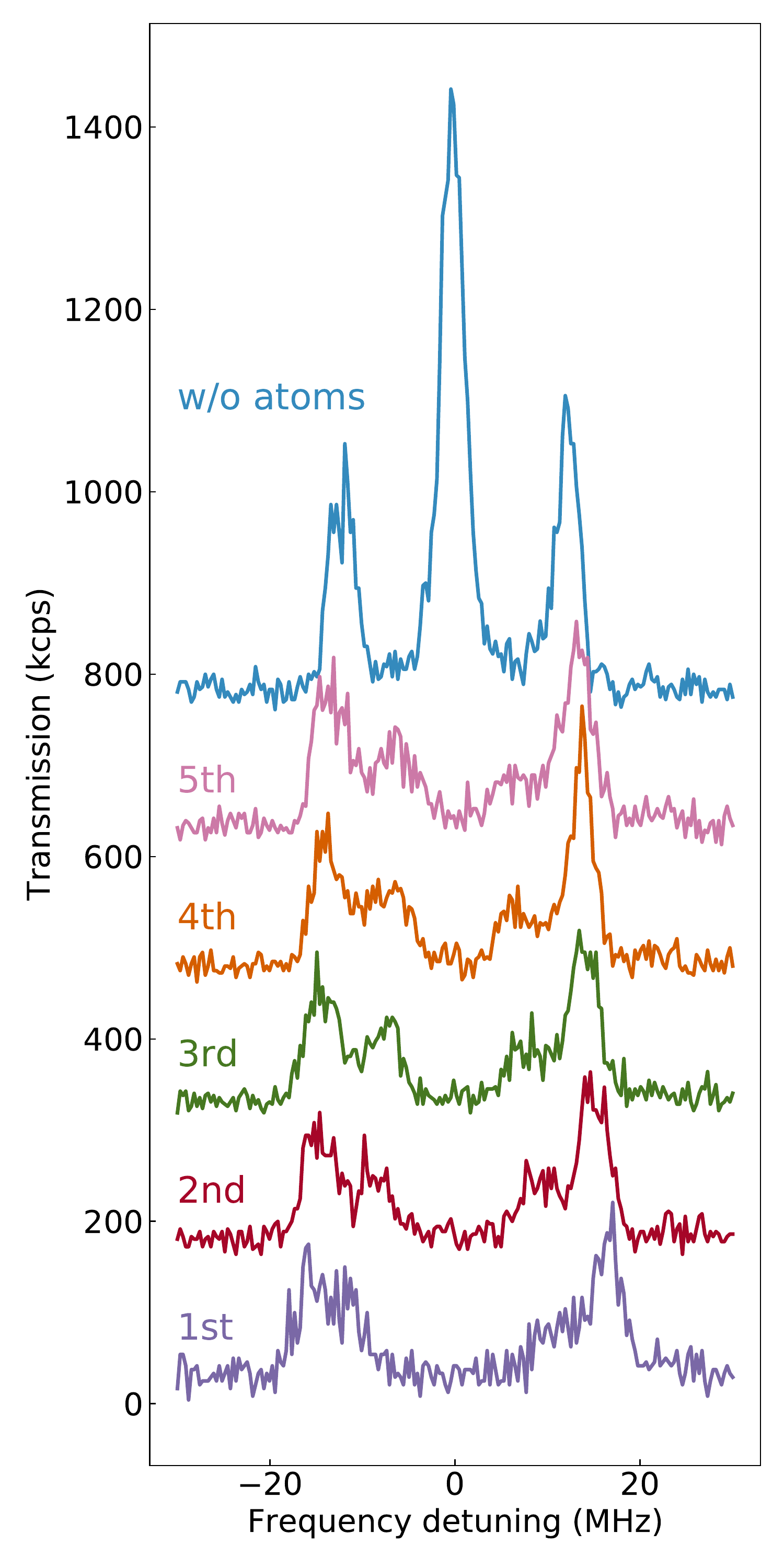}
\caption{Typical transmission spectra for the five consecutive spectroscopy procedures per single MOT loading event.
The experimental condition is the same as that of Fig. 2d in the main text.
The spectra include vertical offsets for clarity.}
\label{sfig-3}
\end{figure*}

The number of the atoms in the MOTs saturates within the loading time of about 500~ms.
After the preparation of the atoms in the MOT, we adjust the resonance frequency of the cavity array close to the F=4$\rightarrow$F'=5 transition.
We input a monitor beam with the power of 130~pW and the frequency locked to the F=4$\rightarrow$F'=5 transition, while changing the cavity lengths of the cavity array (cavity 1, 2, and the connecting fiber), and the transmission of the monitor beam is used as a start trigger for the following sequence.
The MOTs are kept during the monitoring stage, and the positions have a small offset from the nanofibers to keep the cavities without atoms.
When the transmission surpass a threshold, we switch off the beam for the monitor, and change the detuning and intensity of the cooling laser to 0.5$I_s$ and -3.45$\Gamma$, respectively, and hold for 32\ ms for further cooling and loading the atoms into the optical trap.
During the cooling stage, the positions of the atoms are overlapped onto the nanofibers to load them in the traps.
After the loading atoms into the traps, we send a probe pulse and sweep the frequency detuning to the F=4$\rightarrow$F'=5 over $\pm$30\ MHz within 4\ ms.
To observe the transmission spectrum without atoms (empty cavity), we optically pump the atoms into the F=3 state, which is off-resonant to the cavity field, by using the optical pumping pulse irradiated from the side of nanofibers.
We use the F=4$\rightarrow$F'=3 transition for the pumping, and the pulse duration is 1~ms.
After the optical pumping, we send a probe pulse again, and sweep the frequency detuning in the same manner as the previous probe pulse.
We then switch on the cooling and repumping lasers to cool and load the atoms into the optical trap again. 
The cooling duration is 4\ ms.
The spectroscopy procedure repeats five times per single MOT loading event.
Figure~\ \ref{sfig-3} shows typical transmission spectra for a sequence of five spectroscopy procedures, each of which is an average of about 15 sets of data. 
It can be seen that the splittings for the center peak (fiber-dark mode) reduces as the spectroscopy procedure is repeated, which is due to the decrease of the number of the atoms in the optical trap.
We use the average of the second, third, and forth spectra (after taking an average of about 15 sets of data for each of the five spectroscopy procedures in the same way as the spectra in Fig.~\ \ref{sfig-3}) for the data analysis in the main text.

In the data analysis, the observed spectrum without the atom-cavity interaction (empty cavity), where the atoms are optically pumped into the F=3 state, is used to judge whether the cavity array is resonant to the atomic transition or not.
When all cavities (Cavity1, 2 and the channel part) are resonant to the atom, the transmission spectrum has a symmetric triplet peaks centered at the zero atom-probe detuning (see Figs 2a, 3a, 3c in the main text).
In order to choose the data in the relevant condition, we put the following criteria to filter the data; the spectrum of the empty cavity has a largest transmission peak around the atomic resonance frequency within $\pm$0.9 MHz, and has symmetrical sideband peaks at $\pm\sqrt{2}\tilde{v}$ within $\pm$1.8 MHz.

\clearpage

\section*{Acknowledgments}
This work was supported by JST CREST Grant Number JPMJCR1771, JSPS KAKENHI Grant Numbers 16H01055, 18H04293, JST PRESTO Grant Number JPMJPR1662, Japan, and Institute for Advanced Theoretical and Experimental Physics, Waseda University.
S.\,P. acknowledges support from the Marsden Fund of the Royal Society of New Zealand (Contract No. UOA1328).


\end{document}